\documentstyle[12pt]{article}
\begin{document}
\vskip1cm
\def \be{\begin{equation}}
\def \ee{\end{equation}}
\def \l{\label}
\begin{center}
{\bf Discrete-to-continuum transitions and mathematical generalizations in the classical  harmonic oscillator}
\end{center}
\medskip
\begin{center}  Manoelito Martins de Souza\\
Universidade Federal do Esp\'{\i}rito Santo.
Departamento de F\'{\i}sica \\
\par
29060-900  Vit\'oria - ES - Brasil
\footnote{e-mail:manoelit@cce.ufes.br }\\
\end{center}
\par
\medskip
\begin{center}
(PACS: $03.20.+i$  $03.50.-z$)\\
\end{center}
\par
\bigskip
\begin{abstract}
\noindent Discrete interaction models for the classical harmonic oscillator are used for introducing new mathematical generalizations in the usual continuum formalism.  Generalized discrete hyperbolic and trigonometric functions are the exact solutions to generalized (varying frequency) discrete classical harmonic oscillators. Energy and angular momentum conservation and the preservation of the Poisson bracket structure with discrete interactions is discussed.
\end{abstract}
\noindent

\section{Introduction}
\noindent Discrete to continuum transitions in the classical harmonic oscillator have been studied \cite{gr-qc/0301081} with a simple model of a discrete harmonic oscillator where the continuous interaction is replaced by successive exchanges of momentum-energy packets, classical mimics of the interaction quanta. The subject is retaken here with an eye on mathematical generalizations of the usual formalism as a first step towards future connections with a proper quantum formalism. 

Discrete to continuum transitions is a research subject in mathematics, physics and engineering; see, for example \cite{Barker} and the references there in.  
Discretization in physics is an old and very often used tool. Generally, it is a lattice structure \cite{Wegner} taken as an approximation of supposedly continuous physical systems whose continuity is then retrieved by taking a limit with the lattice size going arbitrarily close to zero. In such a context, discretization is just a mathematical artifact for overcoming computational difficulties with  continuous systems.  In contradistinction, the idea behind the naive discrete interaction models considered here is of a genuinely quantum (in the sense of discrete) world. Increments like $\Delta t_{j},\,\Delta {\vec{r}}_{j},\,\Delta {\vec{p}}_{j}$, of time, position and momentum, respectively, although extremely small, are fixed and essentially non-null and can be taken as null only in an approximate sense. Nonetheless, the standard picture of a continuum interaction is still retrieved -- otherwise these discrete models would be completely irrelevant-- as the number of interactions grows larger, $n>>1$. Actually there is an interval 
\be
\l{int}
\frac{1}{(\omega_{n}\Delta t_{n})^2}>>n>>1,
\ee
for the integer $n$ on which the continuum interaction picture is a good effective approximation; outside it the interaction discreteness cannot be ignored.  $\omega_{n}$ is a system characteristic frequency and $\Delta t_{n}$ is the time interval between the $n-th$ and the $(n+1)-th$ interaction. So, in its time evolution a system goes through three well defined phases which may present very distinct characteristics \cite{gr-qc/0301081}.

 That the continuum is retrieved in the limit of zero latticing  in the standard lattice calculation is, of course, no surprising since it was tailored for that.  Remarkable, in the case of the genuine  discrete interactions, is the natural and unavoidable emergence of an effective continuum interaction from fixed increments as their number $n$ grows larger, even though there are no ' '\,forces" in between the exchange of ' '\,quanta", hence no potential energy and no Hamiltonian.
This discrete-to-continuum transition for large values of $n$ reminds the Bohr Correspondence Principle, as discussed in \cite{gr-qc/0301081}. The limit of zero increment is never taken. The continuum here is the approximate picture. Also remarkable is the new (not contained in continuum interaction picture) physics outside the good-effective-approximation range for either $n$ too small  (ultra short distances) or too large (ultra large distances) in the sense of (\ref{int}).  From this point of view, the exact solutions of discretely interacting systems become much more interesting on the light that the standard continuum solutions are incomplete information \cite{Khrennikov}  much in the same way as the quantum solutions are to the classical (the approximate) ones. Discrete or continuum fundamental physical interactions is to be experimentally decided: Discreteness effects for $n$ outside the interval (\ref{int})  can be experimentally observed. There are, perhaps, signs that they have already \cite{gr-qc/0301081} been observed. 
The point in \cite{gr-qc/0301081} was to show the continuum solution of the standard harmonic oscillator emerging as first-order approximation in the large-$n$ limit of solutions of the discrete harmonic oscillator.  Discrete exact solutions of discrete harmonic oscillator, for all values of $n$, are the subject of this paper.
It is structured in the following way. After briefly reviewing the idea of classical discrete interactions in Section II, we treat the time evolution of a discrete harmonic oscillator as a generalized discrete rotation in phase space introducing the generalized discrete trigonometric functions as its exact solutions; discrete hyperbolic harmonic oscillator (an inverted harmonic potential) is also considered.  In Section III we deal with the simplest case of constant time interval and its transition, in the large $n$ limit, to the standard continuous harmonic oscillator.
 Angular momentum and energy conservation and the preservation of the Poison bracket structures are discussed in Section IV where the validity of Eq. (\ref{int}) is established.
 Finally, with the Conclusions in Section V, we discuss the various possible alternative physical interpretations. 
 
\section{Discrete interactions}

A non-relativistic regime, which is appropriate for dealing with harmonic oscillator, is assumed. So, we will consider the relative movement of the reduced mass $m$ of a non-relativistic two body-system. At the initial time $t_{0},$ with initial position ${\vec{r}}_{0}$ and momentum 
${\vec{p}}_{0}$, it freely moves on a rectilinear trajectory until its, after then, first interaction at 
$$t_{1}=t_{0}+\Delta t_{0},$$
$${\vec{r}}_{1}={\vec{r}}_{0}+\Delta{\vec{r}}_{0}={\vec{r}}_{0}+\frac{{\vec{p}}_{0}}{m}\Delta t_{0},$$
that changes its momentum to 
$${\vec{p}}_{1}={\vec{p}}_{0}+\Delta{\vec{p}}_{0}.$$ 
At the $n-th$ interaction,
\be
\l{tn}
t_{n}=t_{0}+\sum_{j=0}^{n-1}\Delta t_{j}.
\ee

\be
\l{vpn}
{\vec{p}}_{n}={\vec{p}}_{n-1}+\Delta{\vec{p}}_{n-1}={\vec{p}}_{0}+\sum_{j=0}^{n-1}\Delta{\vec{p}}_{j},
\ee
\be
\l{vrn}
{\vec{r}}_{n}={\vec{r}}_{n-1}+\Delta{\vec{r}}_{n-1}={\vec{r}}_{n-1}+\frac{{\vec{p}}_{n-1}}{m}\Delta t_{n-1}={\vec{r}}_{0}+\frac{1}{m}\sum_{j=0}^{n-1}{\vec{p}}_{j}\Delta t_{j}.
\ee
The position vector, continually changing, describes a polygonal trajectory. There is a sudden change of momentum at each interaction point, a vertex of the polygon. Momentum is clearly a discrete parameter, or at least it changes discretely.
The continuous variables time and position enter in the description of the motion as if they were also discrete parameters only because the interaction events are our reference points for time counting; this is not a lattice calculation.  The system time evolution in phase-space consist of  changes of position produced along a non-null lapse of time but with a fixed momentum, and changes of momentum produced in a null lapse of time but at a fixed position. This is the essence of Eqs. (\ref{tn}-\ref{vrn}), of discrete interactions. Its physical content is that any allowed apparently continuum path in phase space is, actually, just a first order approximation of these discrete change of momentum followed by changes of positions. This is a point of contact with Quantum Field Theories: Discreteness in the interactions through an exchange of (quantum) discrete and defined amount of momentum and energy.
One cannot talk of force or acceleration, just of sudden change of momentum or velocity. Force and acceleration, as well as all physical concepts that depend on them, become effective concepts, valid only in appropriate limits, for which the picture of a continuum interaction is a good approximation.  

\subsection{Discrete Oscillator}

The Eqs. (\ref{tn}-\ref{vrn}) are generic relations valid for any non-relativistic discretely interacting system. Let us now make Eq. (\ref{vpn}) specific to a discrete harmonic oscillator DHO with the following phenomenological assumption about the interaction
\be
\l{hooke1}
\Delta{\vec{p}}_{n-1}=-m\,\omega^{2}\,{\vec{r}}_{n-1}\Delta t_{n-1},
\ee
which can be taken as a definition of a DHO. Eqs. (\ref{vrn},\ref{hooke1}) restrict the DHO to   the plane determined by
${\vec{r}}_{0}$ and ${\vec{p}}_{0}$, as in any central force motion. These evolution equations can be put in a very convenient matrix form
\be
\l{mat}
\pmatrix{{\vec{r}}_{n}\cr {\vec{p}}_{n}\cr}=\pmatrix{1&\Delta t_{n-1}/m\cr -m\omega^2\Delta t_{n-1}& 1\cr}\pmatrix{{\vec{r}}_{n-1}\cr {\vec{p}}_{n-1}\cr}.
\ee
For a dimensional homogenization in the phase space we make a change of variable
\be
\l{unit}
{\vec{r'}}_{n}=\sqrt{m\,\omega}\,{\vec{r}}_{n}\qquad {\vec{p'}}_{n}={\vec{p}}_{n}/\sqrt{m\,\omega},
\ee
so that Eq. (\ref{mat}), with $\theta_{n-1}=\omega\Delta t_{n-1}$, becomes
\be
\l{mat1}
\pmatrix{{\vec{r'}}_{n}\cr {\vec{p'}}_{n}\cr}=\pmatrix{1&\theta_{n-1}\cr -\theta_{n-1}& 1\cr}\pmatrix{{\vec{r'}}_{n-1}\cr {\vec{p'}}_{n-1}\cr}:=G(\theta_{n-1})\pmatrix{{\vec{r'}}_{n-1}\cr {\vec{p'}}_{n-1}\cr},
\ee
where we have introduced the finite discrete transformation $G(\theta_{n-1}).$ From now on we will drop the primes on the new variables.
 Then
\be
\l{mat2}
\pmatrix{{\vec{r}}_{n}\cr {\vec{p}}_{n}\cr}=\prod_{j=0}^{n-1}G(\theta_{j})\pmatrix{{\vec{r}}_{0}\cr {\vec{p}}_{0}\cr}.
\ee
$G(\theta_{j})$ is a quasi-unimodular transformation
\be
\l{dG}
\det G(\theta_{j})=1+\theta_{j}^2\approx1,
\ee
since $\theta_{j}<<1$,
\be
\l{Gm1}
 G(\theta_{j})^{-1}=(1+\theta_{j}^2)^{-1}G(-\theta_{j}).
\ee
The semblance of $G(\theta_{j})$ of Eq. (\ref{mat1}) with an infinitesimal (continuous) rotation suggests the introduction of generalized discrete trigonometric functions
\be
\l{mat3}
\pmatrix{\,{\hbox{cosg}}\theta(n) & {\hbox{sing}}\theta(n)\cr -{\hbox{sing}}\theta(n)& {\hbox{cosg}}\theta(n) \cr}:=\prod_{j=0}^{n-1}G(\theta_{j}),
\ee
so that 
\be
\l{mat4}
\pmatrix{{\vec{r}}_{n}\cr {\vec{p}}_{n}\cr}=\pmatrix{\,{\hbox{cosg}}\theta(n) & {\hbox{sing}}\theta(n)\cr -{\hbox{sing}}\theta(n)& {\hbox{cosg}}\theta(n) \cr}\pmatrix{{\vec{r}}_{0}\cr {\vec{p}}_{0}\cr}.
\ee
Therefore,
$$
{\hbox{cosg}}\theta(n)=1-\sum_{j_{1}<j_{2}}^{n-1}\theta_{j_{1}}\theta_{j_{2}}+\sum_{j_{1}<\dots<j_{4}}^{n-1}\theta_{j_{1}}\dots\theta_{j_{4}}-\sum_{j_{1}<\dots<j_{6}}^{n-1}\theta_{j_{1}}\dots\theta_{j_{6}}+\dots
$$
$$
{\hbox{sing}}\theta(n)=\sum_{j}^{n-1}\theta_{j}-\sum_{j_{1}<j_{2}<j_{3}}^{n-1}\theta_{j_{1}}\theta_{j_{2}}\theta_{j_{3}}+\sum_{j_{1}<\dots<j_{5}}^{n-1}\theta_{j_{1}}\dots\theta_{j_{5}}-\sum_{j_{1}<\dots<j_{7}}^{n-1}\theta_{j_{1}}\dots\theta_{j_{7}}+\dots
$$
These generalized discrete trigonometric functions, according to Eq. (\ref{mat4}), constitute the exact solutions of the discrete harmonic oscillator. They 
 are both finite series and can be more compactly expressed with 
\be
\l{T}
\Theta_{n}^{(k)}:=\sum_{0\le j_{1}<\dots<j_{k}}^{n-1}\theta_{j_{1}}\;\theta_{j_{2}}\;\dots\theta_{j_{k}},\qquad0\le k\le n
\ee
$$\Theta_{n}^{(0)}=1,\qquad{\hbox{and}}\qquad\Theta_{n}^{(k)}=0\quad\hbox{if}\quad k>n$$
so that,
\be
\l{cg}
{\hbox{cosg}}\theta(n)\equiv\sum_{s=0}^{[n/2]}(-1)^{s}\Theta_{n}^{\;(2s)},
\ee
\be
\l{sg}
{\hbox{sing}}\theta(n):=\sum_{s=0}^{[n/2]}(-1)^{s}\Theta_{n}^{(2s+1)}, 
\ee
where $[n/2]$ is the largest integer in $n/2$. Or yet, we can just use
\be
\l{cosg1}
{\hbox{cosg}}\theta(n)=\frac{1}{2}(\prod_{j=0}^{n-1}(1+i\theta_{j})+\prod_{j=0}^{n-1}(1-i\theta_{j}))
\ee
\be
\l{sing1}
{\hbox{sing}}\theta(n)=\frac{1}{2i}(\prod_{j=0}^{n-1}(1+i\theta_{j})-\prod_{j=0}^{n-1}(1-i\theta_{j})),
\ee
with $i^2=-1$ and the convention that $$\prod_{j=n_{2}}^{n_{1}}X_{j}:=1\quad\hbox{if}\quad n_{2}>n_{1},\quad\hbox{for any $X_{j}$}.
$$
Then 
\be
\l{Moivre}
{\hbox{cosg}}\theta(n)\pm i\,{\hbox{sing}}\theta(n)=\prod_{j=0}^{n-1}(1\pm i\theta_{j}),
\ee
which is a generalized discrete version of  de Moivre's formula. It leads to
\be
{\hbox{cosg}}\,\theta{(n+1)}\pm i\,{\hbox{sing}}\theta{(n+1)}=(1\pm i\theta_{n})[{\hbox{cosg}}\theta(n)\pm i\,{\hbox{sing}}\theta(n)],
\ee
\be
{\vec{r}}_{n}\pm i\, {\vec{p}}_{n}=[{\hbox{cosg}}\theta(n)\pm i\,{\hbox{sing}}\theta(n)]({\vec{r}}_{0}\pm i {\vec{p}}_{0}),
\ee
making explicit the change-of-phase character of the harmonic oscillator motion in phase space.

Table I lists the generalized discrete trigonometric functions for the first five values of $n$.
 \bigskip
\begin{table}
\caption{The generalized discrete trigonometric functions for the first five values of $n$}
 \bigskip
\begin{tabular}{|l||l|}\hline

\medskip

 \centering
${\hbox{cosg}}\theta(0)=1$ & ${\hbox{sing}}\theta(0)=0$ \\
${\hbox{cosg}}\theta(1)=1$   & ${\hbox{sing}}\theta(1)=\theta_{0}$ \\
${\hbox{cosg}}\theta(2)=1-\theta_{1}\theta_{0}$   & ${\hbox{sing}}\theta(2)=\theta_{0}+\theta_{1}$ \\
${\hbox{cosg}}\theta(3)=1-(\theta_{1}\theta_{0}+\theta_{0}\theta_{2}+\theta_{1}\theta_{2})$    & ${\hbox{sing}}\theta(3)=\theta_{0}+\theta_{1}+\theta_{2}-\theta_{0}\theta_{1}\theta_{2}$  \\
${\hbox{cosg}}\theta(4)=1-(\theta_{0}(\theta_{1}+\theta_{2}+\theta_{3})+\theta_{1}(\theta_{2}+$    & ${\hbox{sing}}\theta(4)=\theta_{0}+\theta_{1}+\theta_{2}+\theta_{3}\,-$  \\ \qquad\quad$+\theta_{3})+\theta_{2}\theta_{3})++\theta_{0}\theta_{1}\theta_{2}\theta_{3}$&$-\,\theta_{0}(\theta_{1}\theta_{2}+\theta_{1}\theta_{3}+\theta_{2}\theta_{3})-\theta_{1}\theta_{2}\theta_{3}$\\\hline

\end{tabular}
\end{table}

\bigskip

\subsection{Some properties}
\noindent This notation for the generalized discrete trigonometric functions, introduced in Eq. (\ref{mat1}), explores the parallelism with the continuous formalism, like in the symmetry
$${\hbox{cosg}}(-\theta(n_{2}))={\hbox{cosg}}\theta(n_{2})$$
$${\hbox{sing}}(-\theta(n_{2}))=-{\hbox{sing}}\theta(n_{2}),$$
which corresponds to the time reversibility of discrete interactions, as $\theta_{j}=\omega\Delta t_{j},$
but some care must be taken. The generalized discrete trigonometric functions are not the truncated series of the standard continuous trigonometric functions. They differ in two aspects. Besides being finite, these series are defined by the products of, in general, distinct elementary transformations. But even when they are equal, as in the particular case discussed in the following section, they are not the partial sums of trigonometric series which, by the way, are their first order asymptotic limits, as we will see. 

The Eqs. (\ref{dG},\ref{Gm1},\ref{mat3}) imply on 
\be
\l{c2ps2g}
{\hbox{cosg}}^{2}\;\theta(n)+{\hbox{sing}}^{2}\;\theta(n)=\prod_{j=0}^{n-1}(1+\theta_{j}^2).
\ee
\be
\l{mat5}
{\big(}\prod_{j=0}^{n-1}G(\theta_{j})){\big)}^{-1}=\prod_{j=0}^{n-1}G^{-1}(\theta_{j})=
\prod_{j=0}^{n}(1+\theta_{j}^2)^{-1}\pmatrix{{\hbox{cosg}}\theta(n) & -{\hbox{sing}}\theta(n)\cr {\hbox{sing}}\theta(n)& {\hbox{cosg}}\theta(n) \cr}
\ee
and 
$$
\prod_{j=n_{1}}^{n_{2}-1}G(\theta_{j})=\prod_{j_{1}=0}^{n_{2}-1}G(\theta_{j_{1}})\prod_{j_{2}=0}^{n_{1}-1}G^{-1}(\theta_{j_{2}})=
$$
\be
\prod_{j=0}^{n_{1}-1}(1+\theta_{j}^{2})^{-1}
\pmatrix{{\hbox{cosg}}\theta(n_{2}) & {\hbox{sing}}\theta(n_{2})\cr -{\hbox{sing}}\theta(n_{2})& {\hbox{cosg}}\theta(n_{2}) \cr}\pmatrix{{\hbox{cosg}}\theta(n_{1}) & -{\hbox{sing}}\theta(n_{1})\cr {\hbox{sing}}\theta(n_{1})& {\hbox{cosg}}\theta(n_{1}) \cr}.
\ee
This parallelism with trigonometric functions hides, however, a far more larger complexity.  According to the definitions Eqs. (\ref{cosg1},\ref{sing1}), ${\hbox{cosg}}\theta(n)$ and ${\hbox{sing}}\theta(n)$ are defined by the set $\{\theta_{0}, \theta_{1},\dots,\theta_{n-1}\}$. So, in principle, there is a function $\theta(n)$, still to be properly defined, but it is already clear that 
\be
\l{n1pmn2}
\theta(n_{1})\pm\theta(n_{2})\ne\theta(n_{1}\pm n_{2}),
\ee
as $\theta(n_{1})$ and $\theta(n_{2})$ are defined by two distinct sets, both starting with $\theta_{0}$ but ending, respectively with $\theta_{n_{1}-1}$ and $\theta_{n_{2}-1}$. In order to further explore the parallelism with the standard continuous trigonometric functions let us {\it{define}},
\be
\l{define}
{\hbox{cosg}}[\theta(n_{1})\mp\theta(n_{2})]:={\hbox{cosg}}\theta(n_{1})\,{\hbox{cosg}}\theta(n_{2})\pm {\hbox{sing}}\theta(n_{1})\, {\hbox{sing}}\theta(n_{2}),
\ee
and then, making use of Eqs. (\ref{cosg1}) and (\ref{sing1}),  get
\be
\l{n1mn2}
{\hbox{cosg}}[\theta(n_{1})-\theta(n_{2}-1)]=\frac{1}{2}[\prod_{j=n_{1}}^{n_{2}-1}(1+i\theta_{j})+\prod_{j=n_{1}}^{n_{2}-1}(1-i\theta_{j}))](\prod_{j_{1}=0}^{n_{1}-1}(1+\theta_{j_{1}}^{2}),
\ee
$$
{\hbox{cosg}}[\theta(n_{1})+\theta(n_{2})]=\frac{1}{2}{\big[}\prod_{j_{1}=0}^{n_{1}-1}(1+i\theta_{j_{1}})^{2}\prod_{j_{2}=n_{1}}^{n_{2}-1}(1+i\theta_{j_{2}})+ \prod_{j_{1}=0}^{n_{1}-1}(1-i\theta_{j_{1}})^{2}\prod_{j_{2}=n_{1}}^{n_{2}-1}(1-i\theta_{j_{2}}){\big]},
$$
\noindent which reduce, respectively, to Eq. (\ref{c2ps2g}) and to 
$${\hbox{cosg}}2\theta(n)={\hbox{cosg}}^{2}\theta(n)-{\hbox{sing}}^{2}\theta(n)$$
for $n_{2}=n_{1}=n$. Then we can see from Eq. (\ref{n1mn2}) and Table 1 that, as a particular case of Eq. (\ref{n1pmn2}), 
$$\theta(n)-\theta(n)\ne\theta(0)$$  and that
$${\hbox{cosg}}[\theta(n)-\theta(n)]=\prod_{j=0}^{n-1}(1+\theta_{j}^2).$$
These peculiarities make the generalized discrete trigonometric functions are mathematically interesting consequences of the quasi-unimodularity of the evolution function $G(\theta_{j})$, Eq. (\ref{dG}) and of the definition (\ref{define}); they disappear if a normalized $G(\theta_{j})$ is adopted, but this implies on new equations of motion and will be discussed elsewhere. Anyway, these deviations from the standard behavior are negligible (cannot be physically detected) in the effectivelly continuum-interaction interval (\ref{int}). 

\section{Combinatorial trigonometric  functions}

The complete specification of the generalized discrete trigonometric functions above depends on the specification of the DHO model. This is realized when each time increment $\Delta t_{j}$ is given as a known function of $r_{j}$. The simplest case is the one where $\Delta t_{j}$ is just a constant, it does not depend on $j$. In the other more generic cases the $r_{j}$-dependence in $\Delta t_{j}$ may be absorbed in $\omega$ by redefining it into $\omega_{j}$, so that they describe harmonic oscillator with a variable, $r$-dependent, frequency. 
Let us consider the simplest case and make
\be
 \Delta t_{j}=\alpha,
\ee
where $\alpha$ is a small constant, $$\theta_{j}=\theta=\omega\alpha<<1,\qquad G(\theta_{j})=G(\theta).$$ The function $\theta(n)$ just reduces to $n\theta,$ and the real functions in Eq. (\ref{Moivre}), the {\it{generalized discrete trigonometric functions}}, reduce to the, let us name it, {\it{ discrete combinatorial trigonometric functions}}
\be
\l{cpms}
{\hbox{cosc}}\,n\theta\pm i\,{\hbox{sinc}}\,n\theta=(1\pm i\theta)^n=\sum_{j=0}^{n}(\pm i\theta)^{j} {n\choose j},
\ee
\be
\l{cc}
{\hbox{cosc}}\,n\theta\equiv\frac{1}{2}[(1+i\theta)^{n}+(1-i\theta)^{n}]=\sum_{s=0}^{[n/2]}(-1)^{s}\theta^{2s} {n\choose 2s},
\ee
\be
\l{sc}
{\hbox{sinc}}\,n\theta\equiv\frac{1}{2i}[(1+i\theta)^{n}-(1-i\theta)^{n}]=\sum_{s=0}^{[n/2]}(-1)^{s}\theta^{2s+1} {n\choose {2s+1}}.
\ee
So, according to Eq. (\ref{c2ps2g}), they satisfy 
\be
\l{c2ps2}
{\hbox{cosc}}^{2}n\theta+{\hbox{sinc}}^{2}n\theta\equiv(1+\theta^2)^{n}. 
\ee
A similarly generalized discrete Bessel function as a finite combinatorial series is defined in \cite{ciprian}.
Combinatorial trigonometric  functions are simpler than the generalized discrete trigonometric functions and, therefore, closer to the trigonometric functions. As now $\theta(n_{1})+\theta(n_{2})=\theta(n_{1}+n_{2})$, it is not necessary to define ${\hbox{cosg}}[\theta(n_{1})\mp\theta(n_{2})]$ and ${\hbox{sing}}[\theta(n_{1})\mp\theta(n_{2})].$ The following identities are satisfied.
\be
{\hbox{cosc}}\,n_{1}\theta\,{\hbox{cosc}}\,n_{2}\theta-{\hbox{sinc}}\,n_{1}\,\theta\,{\hbox{sinc}}\,n_{2}\theta={\hbox{cosc}}\,(n_{1}+n_{2})\theta.
\ee
\be
{\hbox{sinc}}\,n_{1}\theta\,{\hbox{cosc}}\,n_{2}\theta+{\hbox{cosc}}\,n_{1}\theta\,{\hbox{sinc}}\,n_{2}\theta={\hbox{sinc}}\,(n_{1}+n_{2})\theta.
\ee
\be
{\hbox{cosc}}\,n_{1}\theta\,{\hbox{cosc}}\,n_{2}\theta+{\hbox{sinc}}\,n_{1}\theta\,{\hbox{sinc}}\,n_{2}\theta=(1+\theta^2)^{n_{1}}{\hbox{cosc}}(n_{2}-n_{1})\theta.
\ee
\be
{\hbox{sinc}}\,n_{1}\theta\,{\hbox{cosc}}\,n_{2}\theta-{\hbox{cosc}}\,n_{1}\theta\,{\hbox{sinc}}\,n_{2}\theta=-(1+\theta^2)^{n_{1}}{\hbox{sinc}}\,(n_{2}-n_{1})\theta,
\ee
and they lead, for $n_{2}=0$ and $n_{1}=n$,  to
\be
{\hbox{cosc}}(-n \theta)=(1+\theta^2)^{-n}{\hbox{cosc}}\,n \theta,
\ee
\be
{\hbox{sinc}}(-n \theta)=-(1+\theta^2)^{-n}\,{\hbox{sinc}}\,n\theta,
\ee
which agrees with Eq. (\ref{Gm1}).
Also
\be
{\hbox{tanc}}\,n\theta=\frac{{\hbox{sinc}}\,2n\theta}{{\hbox{cosc}}\,2n\theta+(1+\theta^2)^{n}}.
\ee
These expressions become identical to the ones for the trigonometric functions if the ubiquotus term $(1+\theta^2)^{n}$ is replaced by 1, an anticipative sign of the upper limit in Eq. (\ref{int}) for retrieving the continuum functions, $$(1+\theta^2)^{n}\approx1+n\theta^2\approx1.$$

Table II lists the first five discrete combinatorial trigonometric functions.
 \bigskip
\begin{table}
\caption{The first five discrete combinatorial trigonometric functions for the values of $n$}

\begin{center}
\begin{tabular}{|l||l|}\hline
${\hbox{cosc}}\,0\theta=1$ & ${\hbox{sinc}}\,0\theta=0$ \\
${\hbox{cosc}}\,1\theta=1$ & ${\hbox{sinc}}\,1\theta=\theta$ \\
${\hbox{cosc}}\,2\theta=1-\theta^2$   & ${\hbox{sinc}}\,2\theta=2\theta$ \\
${\hbox{cosc}}\,3\theta=1-3\theta^2$    & ${\hbox{sinc}}\,3\theta=3\theta-\theta^3$  \\
${\hbox{cosc}}\,4\theta=1-6\theta^2+\theta^4$    & ${\hbox{sinc}}\,4\theta=4\theta-4\theta^3$  \\\hline

\end{tabular}
\end{center}
\end{table}
For a discrete hyperbolic harmonic oscillator, one with repulsive interactions
\be
\l{hookelh}
\Delta{\vec{p}}_{n-1}=k{\vec{r}}_{n-1}\Delta t_{n-1},
\ee
which corresponds to an inverted (repulsive) harmonic potential, the solutions are the generalized iscrete hyperbolic functions
\be
\l{coshg1}
{\hbox{coshg}}\theta_{n}=\frac{1}{2}(\prod_{j=0}^{n-1}(1+\theta_{j})+\prod_{j=0}^{n-1}(1-\theta_{j}))
\ee
\be
\l{sinhg1}
{\hbox{sinhg}}\theta_{n}=\frac{1}{2}(\prod_{j=0}^{n-1}(1+\theta_{j})-\prod_{j=0}^{n-1}(1-\theta_{j})),
\ee
that are associated to discrete hyperbolic rotations in phase space
\be
\l{math1}
\pmatrix{{\vec{r}}_{n}\cr {\vec{p}}_{n}\cr}=\pmatrix{1&\theta_{n-1}\cr \theta_{n-1}& 1\cr}\pmatrix{{\vec{r}}_{n-1}\cr {\vec{p}}_{n-1}\cr}=G_{h}(\theta_{n-1})\pmatrix{{\vec{r}}_{n-1}\cr {\vec{p}}_{n-1}\cr},
\ee
\be
\l{mat3h}
\pmatrix{\,{\hbox{coshg}}\theta(n) & {\hbox{sinhg}}\theta(n)\cr -{\hbox{sinhg}}\theta(n)& {\hbox{coshg}}\theta(n)\cr}:=\prod_{j=0}^{n-1}G_{h}(\theta_{j}).
\ee
They satisfy
\be
\l{ch2psh2g}
{\hbox{coshg}}^{2}\;\theta_{n}-{\hbox{sinhg}}^{2}\;\theta_{n}=\prod_{j=0}^{n}(1+\theta_{j}^2).
\ee

\subsection{Combinatorial identities}

Eqs. (\ref{cc},\ref{sc},\ref{c2ps2}) or (\ref{coshg1},\ref{sinhg1},\ref{ch2psh2g}) imply on some useful combinatorial identities. Expanding each term in both sides of Eq. (\ref{c2ps2}),
$$
{\hbox{cosc}}^{2}n\theta=\sum^{[n/2]}_{s,s'=0}(-1)^{s+s'}\theta^{2(s+s')}{n\choose 2s}{n\choose 2s'}=$$
\be 
=\sum^{2[n/2]}_{k=0}\sum^{k}_{t=0}(-1)^{k}\theta^{2k}{n\choose 2(k-t)}{n\choose 2t},
\ee
\be
{\hbox{sinc}}^{2}n\theta=
=\sum^{2[n/2]}_{k=0}\sum^{k}_{t=0}(-1)^{k}\theta^{2(k+1)}{n\choose 2(k-t)+1}{n\choose 2t+1},
\ee
\be
(1+\theta^2)^{n}=\sum_{k=0}^{n}{n\choose k}\theta^{2k},
\ee
and equating the coefficients of the $\theta^{2k}$ terms in  both sides of Eq. (\ref{c2ps2}), we have
\be
\l{Id}
(-1)^{k}(\sum^{k}_{t=0}{n\choose 2(k-t)}{n\choose 2t}-\sum^{k-1}_{t=0}{n\choose 2(k-t)-1}{n\choose 2t+1})= {n\choose k},
\ee
where, there is a clear separation, in left hand side, of even and odd terms, a trace of sines and cosines. It can be put in a more symmetrical form as
\be
\l{Id1}
\sum_{t=-k}^{k}(-1)^t {n\choose{k+t}} {n\choose{k-t}}= {n\choose k}.
\ee

\subsection{Retrieving the continuum limit}

For completeness sake we make here a brief review \cite{gr-qc/0301081} of how the effective continuum limit is reached. For large values of $n$ the following approximation is valid
\be
\l{app}
{n\choose k}\approx\frac{n^k}{k!},\qquad\hbox{for}\quad n>>1,
\ee
and this turns Eq.(\ref{cpms}) into 
\be
\l{cpms1}
{\hbox{cosc}}\,n\theta\pm i\,{\hbox{sinc}}\,n\theta=(1\pm i\theta)^n\approx\sum_{j=0}^{n}(\pm i\theta)^{j}\frac{n^j}{j!},
\ee
and then the Eqs.(\ref{cc}) and (\ref{sc}) become, respectively
\be
\l{ccng1}
{\hbox{cosc}}\,n\theta\approx\sum_{s=0}^{[n/2]}(-1)^{s}(\alpha\omega)^{2s}\frac{n^{2s}}{(2s)!}\approx\cos(\alpha\omega n),
\ee
\be
\l{scng1}
{\hbox{sinc}}\,n\theta\approx\sum_{s=0}^{[n/2]}(-1)^{s}((\alpha\omega))^{2s+1}\frac{n^{2s+1}}{(2s+1)!}\approx\sin(\alpha\omega n).
\ee
The middle terms of Eqs. (\ref{ccng1}) and (\ref{scng1}) are partial sums of the series representing the respective trigonometric functions which are then
 the asymptotic limits of the respective largest contributions (Eq. (\ref{app})) from each term of the combinatoric function series for $n\rightarrow\infty$.  
The continuous variable time enters in the description of the motion as a discrete parameter (as it happens also to the position variable)
\be
t_{n}=t_{0}+n\alpha.
\ee
It is important to underline that the effective standard picture of a continuously interacting harmonic oscillator emerges, in a natural and unavoidable way, as the number $n$ of interactions  grows, irrespective of the fixed values of $\omega$ and of $\alpha=\Delta t$. The smallness of $\Delta t$ is never required. That is what characterizes this discrete-to-continuum transition as being of the kind described by the Bohr Correspondence Principle. Just for the sake of comparison we remind  well known standard limiting procedures like, for example, to replace in Eq. (\ref{cpms}) $\Delta t$ by $\frac{\Delta t}{N}$ which, naturally goes to zero as $N$ grows unboundly:
\be
\l{cpms2}
{\hbox{cosc}}\,\omega\Delta t\pm i\,{\hbox{sinc}}\,\omega\Delta t=(1\pm i\frac{\omega\Delta t}{N})^N\asymp e^{i\omega\Delta t}.
\ee
$N$, in contradistinction to $n$, is not the number of interactions, which is meaningless for continuum interaction; it is just an ancillary mathematical artifact.   
Such limiting procedures with time lapse approaching zero with the growth of $N$, lead to a genuine continuum interaction and the last term in the right hand side of Eq. (\ref{cpms2}) represents an exact, not just an asymptotic, expression. For discrete interactions, in contradistinction, where $\Delta t$ is fixed, it is just a first order approximation, and  there will always be  known lower order residues. The larger is $n$ the better the first order contribution (in the sense of Eq. (\ref{app})) series approach their respective trigonometric functions. As $\theta<<1$, the second order contributions are always negligible face the first order ones, but for large enough values of $n$ they may become physically detectable producing new physical effects not contained in the continuum formalism. 

\section{Discrete angular momentum and  Hamiltonian}

For the continuum harmonic oscillator the angular momentum ${\vec{L}}:={\vec{r}}\times {\vec{p}}$ is time independent, and the generator of infinitesimal rotations. For the discrete harmonic oscillator, with the quasi-unimodular elementary transformations (\ref{dG}), the above standard definition for the angular momentum makes it explicitly time dependent ($\theta=\omega\Delta t$)
\be
\l{Ln}
{\vec{L}}_{n}:={\vec{r}}_{n}\times {\vec{p}}_{n}=(1+\theta_{n-1}^{2}){\vec{L}}_{n-1}=\prod_{j=0}^{n-1}(1+\theta_{j}^{2}){\vec{L}}_{0},
\ee
and, besides, since there is no infinitesimal rotations (only polygonal trajectories) it cannot be their generators.
Similarly, the Hamiltonian, i.e. the kinetic energy plus the potential energy, is also a constant of motion and a generator of infinitesimal translation in time for the continuum harmonic oscillator. For the discrete one, however, there is no potential energy, only kinetic energy, and its time evolution is trivial (free evolution) except on a finite set of interaction points. So, the meaning of the concept of a Hamiltonian in a discrete interaction context must be reassessed. Let us consider the discrete Hamiltonian $$H_{n}:=\omega(\frac{p^{2}_{n}}{2}+ \frac{r^{2}_{n}}{2}),$$
written in terms of the homogenized units of Eq. (\ref{unit}). According to Eqs. (\ref{mat1}), (\ref{mat2}) and (\ref{c2ps2g}),
\be
\l{Hn}
H_{n}=\prod_{j=0}^{n-1}(1+\theta_{j}^2)H_{0},
\ee
and so, $H_{n}$ is also explicitly time-dependent $(\theta_{j}=\omega\Delta t_{j})$. 
Actually, $H_{n}$, and also  ${\vec{L}}_{n}$, would be more properly described as being $n$, rather than $t_{n}$, dependent. They remain constant in the time  lapse $\Delta t_{n}$ between the $n-{th}$ and the $(n+1)-{th}$ interactions. At each interaction they get larger by a factor $1+\theta_{n}^2$. 

The discrete time evolution (\ref{dG}) does not preserve the canonical relations either. The Poisson brackets are not preserved
\be
\l{PBnn}
\{F_{0}({\vec{r}}_{0},{\vec{p}}_{0}),G_{0}({\vec{r}}_{0},{\vec{p}}_{0})\}_{{\vec{r}}_{0},{\vec{p}}_{0}}\not=\{F_{n}({\vec{r}}_{n},{\vec{p}}_{n}),G_{n}({\vec{r}}_{n},{\vec{p}}_{n})\}_{{\vec{r}}_{0},{\vec{p}}_{0}},
\ee
where $F_{n}$ and $G_{n}$ are functions of ${\vec{r}}_{n}$ and ${\vec{p}}_{n}$. For example,
\be
\{p_{n},r_{n}\}=\{p_{0},r_{0}\}\prod_{j=0}^{n-1}(1+\theta_{j}^2).
\ee
 This non preservation of $H_{n},$ ${\vec{L}}_{n}$ and of the Poisson bracket is a consequence of the non-unimodular character (\ref{dG}) of the elementary transformation.

Whereas the condition $n>>1$, as an inferior bound, is required for retrieving the standard continuous trigonometric functions as first order approximations for the above DHO solutions, the preservation of the canonical relations, or equivalently, the conservation of $H_{n}$  (and of ${\vec{L}}_{n}$) imposes the upper bound condition
$$
n\theta^2<<1,
$$
of the relation (\ref{int}) for the validity of continuum interaction as a good first order approximative description. 
Eq. (\ref{Ln}), written as
\be
\frac{{\vec{L}}_{n}-{\vec{L}}_{n-1}}{\Delta t_{n-1}}=\omega^{2}\Delta t_{n-1}{\vec{L}}_{n-1}
\ee
shows that if $\Delta t_{n-1}$ could be made infinitely close to zero we could have a time derivative (if the ratio on the left hand side remains well defined) and ${\vec{L}}_{n}$ would be conserved. The same goes for $H_{n}$.
But for the discrete oscillator $\Delta t_{j}$ is fixed, it cannot go to zero.   The emerging 
 potential energy, as an effective concept, corresponds to the energy of the still travelling (not yet absorbed) exchanged quanta.

\section{Physical interpretations and Conclusions}

It is plausible that in our quotidian  $\Delta t_{j}$, the time interval between two consecutive discrete interactions, being extremely small, $\omega\Delta t_{n}<<1$, be beyond any today{´}s conceivable experimental detection. Therefore, even if there is no legitimate infinitesimal transformation but only finite ones, in a practical sense of approximation, at the measure of $\omega\Delta t_{n}<<1$, they can be treated as if they were infinitesimal \cite{gr-qc/0301081}. 
Rigorously $H_{n}$ and ${\vec{L}}_{n},$ with quasi-unimodular elementary transformations, are not time-independent, but just quasi. $$H_{n}=(1+\theta_{j}^2)H_{n-1}\approx H_{n-1}$$
$${\vec{L}}_{n}=(1+\theta_{n-1}^{2}){\vec{L}}_{n-1}\approx{\vec{L}}_{n-1}.$$

Up to a first order approximation both, $H_{n}$ and ${\vec{L}}_{n}$ appear to be time independent and the system can be taken as a continuum one.

We list, after all these considerations, three possible alternative scenarios:
\begin{enumerate}
\item The time evolution equation (\ref{mat1}) is correct but Eqs. (\ref{Ln}) and (\ref{Hn}) are not the correct expressions for, respectively the angular momentum and the energy of a DHO. They should have a factor $1/\sqrt{\prod_{j=0}^{n-1}(1+\theta_{j}^2)}$ that would make both $H_{n}$ and ${\vec{L}}_{n},$ time independent but still could not preserve the canonical relations. The preservation of the canonical relations with all its implications (vide quantum mechanics, for instance) would be just a first order approximation artifact and not a fundamental truth of nature. This would be, of course, a very controversial interpretation;
\item The equations (\ref{mat1}), (\ref{Ln}) and (\ref{Hn}) are correct. The explicit time dependence of Eqs. (\ref{Ln}) and (\ref{Hn}) shows that they do not describe a closed system.  An energy and angular momentum supplier is missing in this description.  On the other hand, all the terms in these equations are definite positive which means that is it not possible to build a closed  (energy and angular momentum conserving) system only with  discrete classical harmonic interactions. The concept of energy-and-angular-momentum conserving classical harmonic oscillator, rigorously, does not exist in a discrete interaction context; it is just another asymptotic physical idealization like many others (ideal gas, reversible process, etc.) in physics.
\item  Equation (\ref{mat1}) is not correct; it needs to be normalized into
\be
\l{nmat1}
\pmatrix{{\vec{r}}_{n}\cr {\vec{p}}_{n}\cr}=\frac{1}{\sqrt{1+\theta_{n-1}^2}}\pmatrix{1&\theta_{n-1}\cr -\theta_{n-1}& 1\cr}\pmatrix{{\vec{r}}_{n-1}\cr {\vec{p}}_{n-1}\cr}.
\ee
Then the new elementary transformation
\be
\l{nG}
G(\theta_{n-1})=\frac{1}{\sqrt{1+\theta_{n-1}^2}}\pmatrix{1&\theta_{n-1}\cr -\theta_{n-1}& 1\cr}
\ee
is unimodular, leads to normalized generalized discrete trigonometric functions, with
\be
\l{nc2ps2g}
{\hbox{cosg}}^{2}\;\theta(n)+{\hbox{sing}}^{2}\;\theta(n)=1,
\ee
replacing (\ref{c2ps2g}), and so, preserves the Hamiltoninan, the angular momentum and the Poisson Bracket structure. We do, certainly, favor this alternative but it leads to new second order (on $\theta$) effects that, in principle, can and should be experimentally checked. Its discussion will be presented elsewhere in a quantum mechanics context.
\end{enumerate}
A fourth alternative, of course, could be quantum exchange being just an approximation for an actually continuously interacting world. As it is contradicted by all the great theoretical and experimental achievements of quantum theory it was not even listed.
In summary, the possible idea of a fundamentally discrete interaction puts in check the entire structure of physics based on interaction continuity ---is it exact or just an approximation experimentally validated by the smallness of the interaction quanta?

\end{document}